# Single-shot fringe projection profilometry based on Deep Learning and Computer Graphics

FANZHOU WANG,[1,2,†] CHENXING WANG,[1,2,†,*] AND QINGZE GUAN [3]

[1]School of Automation, Southeast University, 2# Sipailou, Xuanwu, Nanjing, 210096, China
[2] Key laboratory of Measurement and Control of Complex Systems of Engineering, Ministry of Education, Southeast University, Nanjing, 210096, China
[3]Department of Computer Science, University College London, Gower St, Bloomsbury, London, WC1E 6BT, United Kingdom
†Authors contributed equally to this work.
*cxwang@seu.edu.cn

**Abstract:** Multiple works have applied deep learning to fringe projection profilometry (FPP) in recent years. However, to obtain a large amount of data from actual systems for training is still a tricky problem, and moreover, the network design and optimization still worth exploring. In this paper, we introduce computer graphics to build virtual FPP systems in order to generate the desired datasets conveniently and simply. The way of constructing a virtual FPP system is described in detail firstly, and then some key factors to set the virtual FPP system much close to the reality are analyzed. With the aim of accurately estimating the depth image from only one fringe image, we also design a new loss function to enhance the quality of the overall and detailed information restored. And two representative networks, U-Net and pix2pix, are compared in multiple aspects. The real experiments prove the good accuracy and generalization of the network trained by the data from our virtual systems and the designed loss, implying the potential of our method for applications.

## 1. Introduction

Fringe projection profilometry (FPP) is a classic solution for 3D shape scanning. It projects coded fringes to an object and captures the deformed fringe images modulated by the object's surface. Then the 3D shape is reconstructed by demodulating the fringe signals, and the 3D point cloud is obtained through further calibration algorithms. Although FPP has been applied to multiple scenarios [1-4], it still faces the difficulty to balance the accuracy and the speed. N-step phase-shifting algorithm [5] is precise and commonly used, but it requires projecting and capturing more than three different fringe images. This process is time-consuming for dynamic measurements. With a single-shot fringe image, Fourier-transform profilometry (FTP) [6] extracts the carrier spectra for 3D shape reconstruction. Unfortunately, its accuracy would be affected by spectra overlaps when processing complex shapes. To solve this problem, windowed Fourier transform [7], wavelet transform [8] such like spectra analysis methods are introduced, but they need heavy calculations and presetting parameters. In addition, the methods above usually obtain the phase wrapped into a range of 2π, thus unwrapping algorithms are needed to further restore the true 3D shapes. Similarly, temporal phase unwrapping methods, such as gray-code method [9,10] and multi-frequency method [11,12], are simple to calculate, but they require projecting additional multiple fringe images. Spatial phase unwrapping methods, such as branch-cut method [13], flood method [14], and Laplacian operator method [15], can perform unwrapping from a single image, but they need massive calculation and the accuracy is sensitive to the noise, shadow, or height jump.

In recent years, deep learning presents powerful performance with the improvement of the neural network structure and the computing power. Plenty of studies have proved that deep learning performs superior to traditional algorithms in terms of speed and robustness, which are used for fringe denoising [16-18], fringe analysis [19,20], and phase unwrapping [21-23]. However, these works conduct experiments with limited datasets and only focus on a single step of the FPP system, which means that multiple networks must be integrated to construct a complete system. Naturally, the training process and the preparation of training sets are troublesome for integrating networks, and such integration inevitably accumulates errors. Refreshingly, some researches [24-26] directly map a single fringe image to its height/depth image with a single network, and some of them explore to generate training data by simulating with mathematical expressions [24] or by building a digital virtual twin [26]. These explorations are effective and worth being developed for generalization applications of FPP.

Existing works used in FPP basically choose the convolutional neural networks (CNNs). For instance, an optical fringe pattern de-noising convolutional neural network (FPD-CNN) model is proposed in [18]; a CNN model for wrapped-phase calculation is proposed in [20]; the U-Net [27] is improved for phase-unwrapping in [23]. To guide the selection of a suitable network for the single-shot FPP, a comparison of three CNNs is conducted in [25], including fully convolutional networks (FCN) [28], Autoencoder networks (AEN) [29], and U-Net [27], and U-Net is concluded performing the best due to its symmetric structure and feature map concatenation. In fact, except for CNN, other network models also emerge with powerful performance recently, such as the adversarial generative network (GAN) [30], which generates data in a certain style through an adversarial procedure. Among GANs, pix2pix [32] establishes the conversion from an image to an image and shows excellent performance in generating images with details.

In this paper, we also select the strategy of mapping a fringe image to a depth image to reconstruct the 3D shape efficiently. We introduce the computer graphics to simulate a virtual FPP system that conveniently generates large-scale and diverse data samples as requirements. The methods to construct the virtual system are given in detail. Furthermore, variable factors influencing the accuracy of a real FPP system are researched thoroughly. With different combinations of these parameters being set, our simulated virtual FPP system renders different large datasets that are investigated and evaluated their influence on the accuracy and generalization of the network. In addition, we design a new loss function considering the structure similarity of objects and the detail information, which is proved effective for improving the overall and detailed accuracy of the generated depth image. U-Net and pix2pix, the representatives of CNNs and GANs respectively, are compared by multiple experiments to explore the better solution for estimating the depth image from a single-shot fringe image. Finally, the real experiment further verifies the accuracy and the generalization ability of our method.

## 2. The construction of a virtual FPP system and the rendering of datasets

Sufficient training data are the guarantee of excellent performance for deep learning networks, which, however, is a tricky problem for most deep learning methods used for FPP. Recently, computer graphics has been successfully introduced for dataset generation [33-35], inspiring us to construct a virtual FPP system to establish diverse datasets conveniently. This section explains the details of constructing the virtual FPP system and rendering data samples.

### *2.1 The selection of 3D models*

The virtual objects used in the virtual FPP system can be selected from existing 3D model datasets, such as ModelNet [36], ShapeNet [37], ABC [38], Thingi10K [39], etc. Considering the effective working distance of FPP in visible light (within 1~2m), we select the Thingi10K dataset that contains various 3D models of common objects, such as sculptures, vases, and dolls, as shown in Fig. 1. The variety and the magnitude of these models help to generate large-scale and diverse data samples as needed.

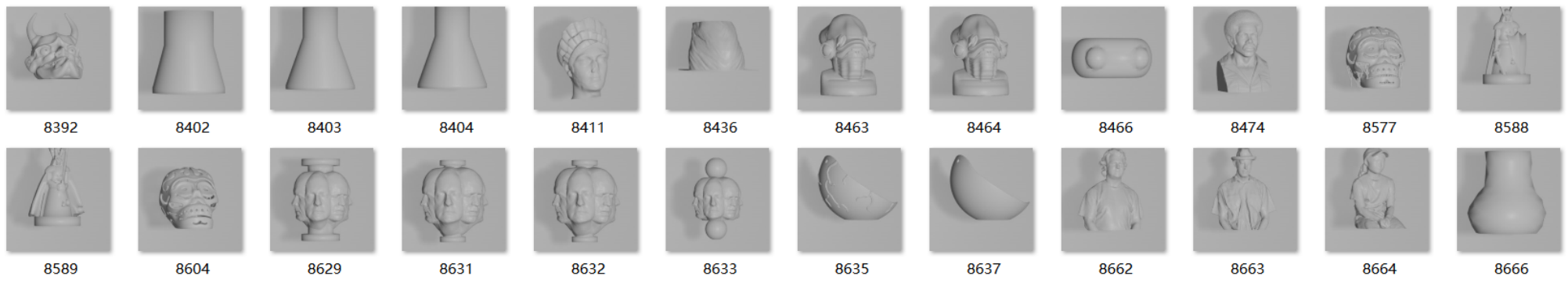

Fig. 1. Some models from Thingi10K.

## *2.2. The construction of a virtual FPP system*

Computer graphics is good at presenting the real-world scene in a virtual form. Among various graphics software, Blender is an open-source 3D creation suite, which is powerful and can generate images by Python in batch. In Blender, a virtual camera and a virtual projector can be placed in the "Layout", as shown in Fig. 2(a) and 2(b). The virtual system works the same as a real FPP system, i.e., the projector projects sinusoidal fringes onto an object, and the deformed fringes are captured by a camera. Blender renders fringe images by setting the compositing node "Render Layers" to "Image", and renders depth images by setting it to "Depth" (shown in Fig. 2(c)).

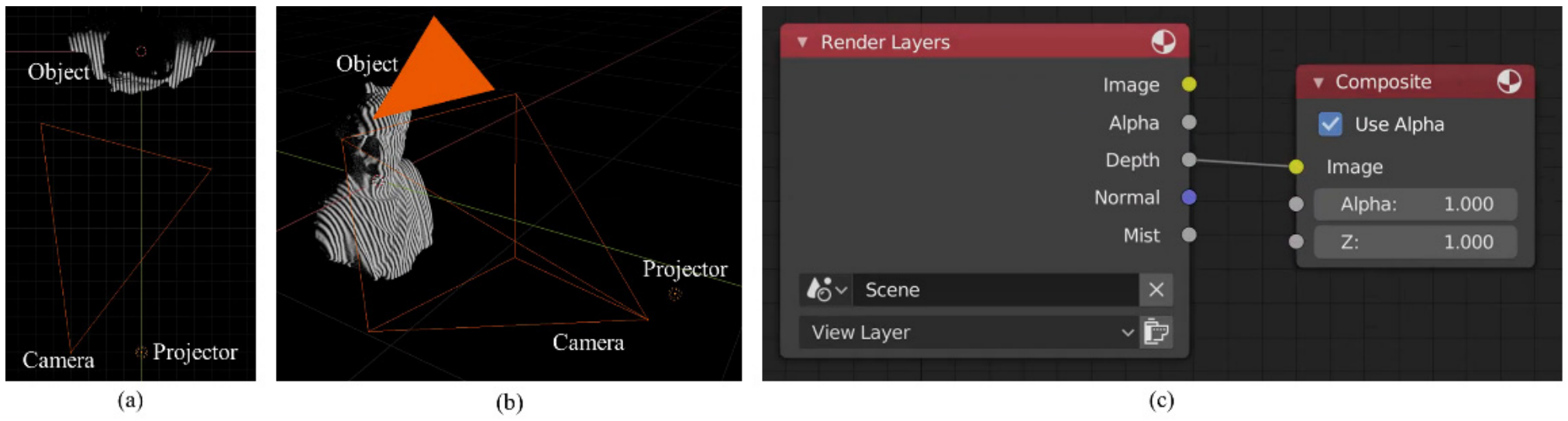

Fig. 2. (a) Aerial view of the scene layout in Blender; (b) side view of the scene layout in Blender; (c) the compositing node tree of this blender system.

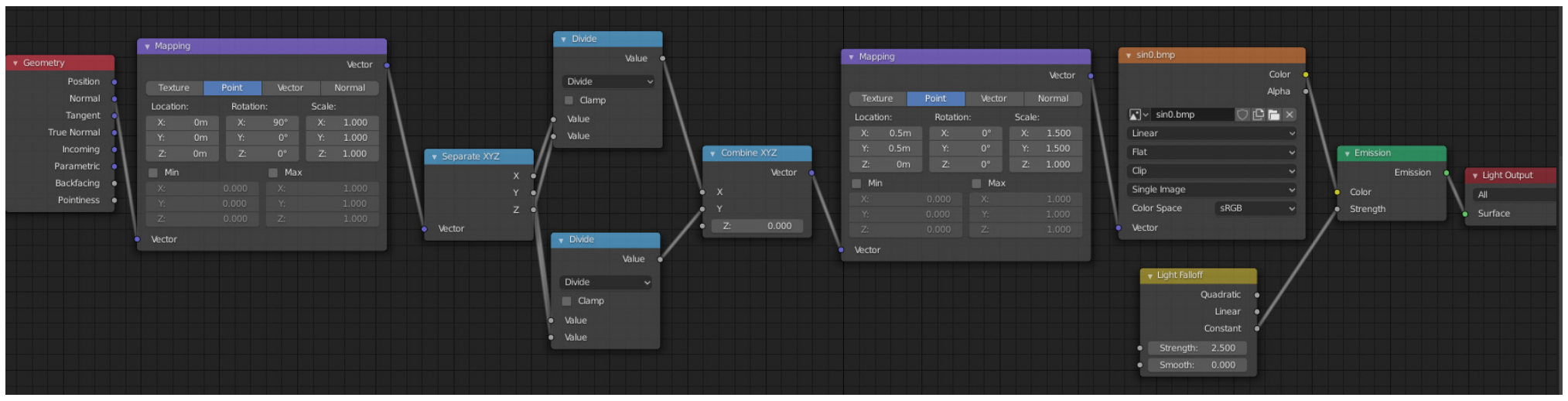

Fig. 3. The shading node tree of constructing a projector.

Some elements of our virtual FPP system in Blender include:

1) **Camera:** the type is set to "perspective", and its position and rotation angle can be adjusted;
2) **Projector:** it is modeled as a point light source, and its shading node tree is designed as Fig. 3 to project parallel sinusoidal fringes, where each node is explained in Appendix A;
3) **Objects:** 3D models are loaded in and are scaled to a proper size;
4) **Background:** some indoor environment maps can be imported into the "World" setting (the effect is shown in Fig. 4), and the shading node tree of "World" is set as Fig. 5, where the rotation angle and the brightness of the background can be randomly changed;

5) **Rendering:** the rendering engine is set as the physically-based path tracer "Cycles", and the sampling integrator is set as "Branched path tracing".

6) **File format:** any common image format is permitted for fringe images, but the depth images should be saved in Open_EXR format to retain the original depth information.

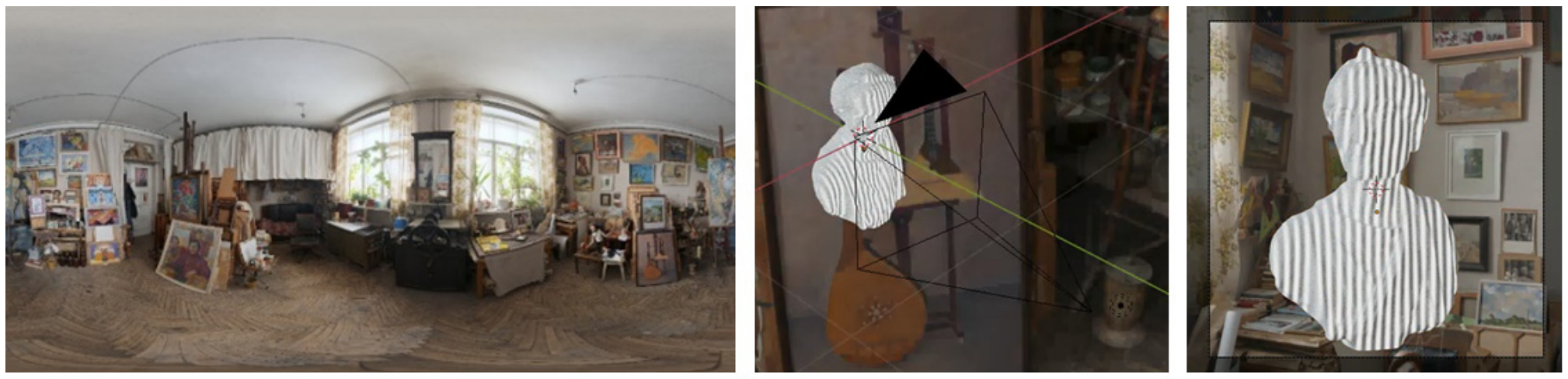

(a) (b) (c)

Fig. 4. (a) An HDRI environment map [40]; (b) the side view of the scene layout after importing an environment map; (c) the camera view after importing an environment map.

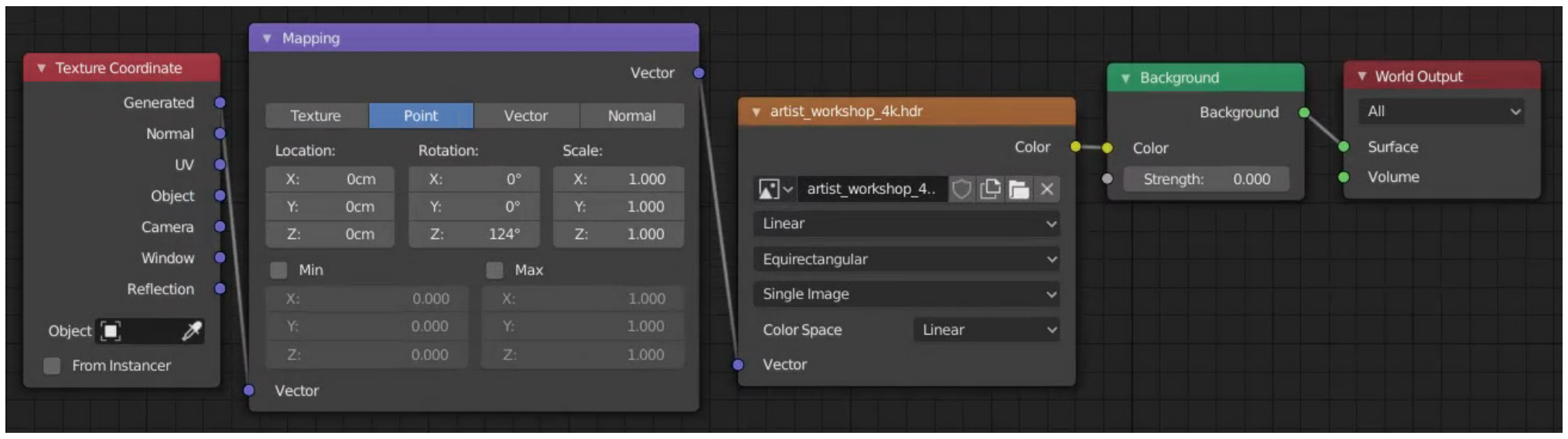

Fig. 5. The shading node tree of "World" [41].

## *2.3 The factors enhancing the reality of the virtual FPP system*

To enhance the generalization of our network, not only the 3D models to construct the training set should be rich and diverse, the settings of the virtual system also have to be adjusted to accord with various possible measurement environments. The input of our network is a sinusoidal fringe image with a usual mathematical description as

$$I(x,y)=a(x,y)\cos[2\pi fx+\varphi(x,y)]+b(x,y)+n(x,y), \quad (1)$$

where $a(x, y)$ is the amplitude intensity, $f$ is the frequency deciding the fringe period, $\varphi(x, y)$ is the phase describing the shape of an object, $b(x, y)$ is the background, and $n(x, y)$ is the noise. These parameters are changed in different measurements, thus the main factors influencing these parameters in practice should be taken into consideration in the virtual system setting, and they are analyzed as follows.

**The period of fringes**

According to the classical calibration theory [42], the image coordinate $[x, y]$ is described as

$$\begin{cases} x=f_x\dfrac{X}{Z}+c_x \\ y=f_y\dfrac{Y}{Z}+c_y \end{cases}, \quad (2)$$

where $[f_x, f_y]$ are the focal lengths of the camera, $[X, Y, Z]$ denote the camera(projector) coordinate and $[c_x, c_y]$ are the optical centers of the camera(projector).

In practice, optical centers vary from different devices, and the focal lengths vary as the random position of objects or the changed depth of objects' surfaces. The difference of optical centers leads to different imaging locations for the object in an image, which will not cause errors for the task of this paper. However, the changes of focal lengths would cause the fringe period of the captured fringe image to be zoomed, which may correspond to the change of $f$ or $\varphi(x, y)$ in Eq.(1). This type of change influences the depth extraction and so is necessary to be considered. In the virtual FPP system, this can be simulated by setting various periods of the projected fringes (adjusting the "scale[0]" in the 2nd "mapping" node in Fig. 3).

**The pose between the camera and the projector**

The space geometry relation between the camera coordinate $[X_c, Y_c, Z_c]$ and the projector coordinate $[X_p, Y_p, Z_p]$ can be described as [42]

$$\begin{bmatrix} X_c \\ Y_c \\ Z_c \end{bmatrix} = R \begin{bmatrix} X_p \\ Y_p \\ Z_p \end{bmatrix} + t. \tag{3}$$

where $R$ and $t$ represent the rotation and the translation matrixes, respectively. To simulate this position relationship, we rotate the projected fringes around the optical axis of the projector (by adjusting the "rotation[2]" in the 2nd "mapping" node in Fig. 3) and set different angles between the camera and the projector to simulate different $R$ and $t$.

**The amplitude intensity and background**

The amplitude intensity of a fringe image, corresponding to $a(x, y)$ in Eq.(1), is generally decided by the material/texture of objects, the power of the projector, and the brightness of the background, which can be set conveniently in the virtual FPP system (by adjusting the "strength" in the "background" node in Fig. 5). And the environment map can be shifted or rotated multiple times to simulate that the objects are located in different backgrounds (by adjusting the "rotation[2]" in the "mapping" node in Fig. 5).

With the factors analyzed above, the virtual FPP system is set to generate the data much close to the ones from a practical system, which thus helps to improve the practicability of the trained network.

## 3. The Networks and the designed loss function

U-Net has been proved the best compared with some other CNNs [25] used in the FPP system. GAN reveals powerful ability in image generation, so it is explored whether it performs better on the depth prediction in this paper. Below the architectures of the U-Net and a conditional GAN (cGAN) named pix2pix are introduced simply, and the design of the loss function for better details and accuracy is also explained.

### *3.1 The Network Architecture*

**U-Net**

U-Net [27] follows an encoder-decoder structure. The encoder down-samples the input images to extract features, and the decoder up-samples the feature maps to obtain a high-resolution output image. U-Net also has a special structure of skip-connection so that larger-scale feature maps can be directly sent to the up-sampling process, and therefore, the output process and input process share the low-level information. Based on these structures, U-Net learns with less data but achieves higher precision. As U-Net is a part of pix2pix, its structure is given as follows.

**pix2pix**

Pix2pix [32] contains a generator and a discriminator. The generator produces fake images, and the discriminator tries to identify the fake ones, guiding the generator to produce a fake image much closer to the target output. Fig. 6 presents the architecture of pix2pix. The network of pix2pix is shown in Fig. 7, where the generator has a U-Net shape and the discriminator is "patchGAN", a multi-layer CNN.

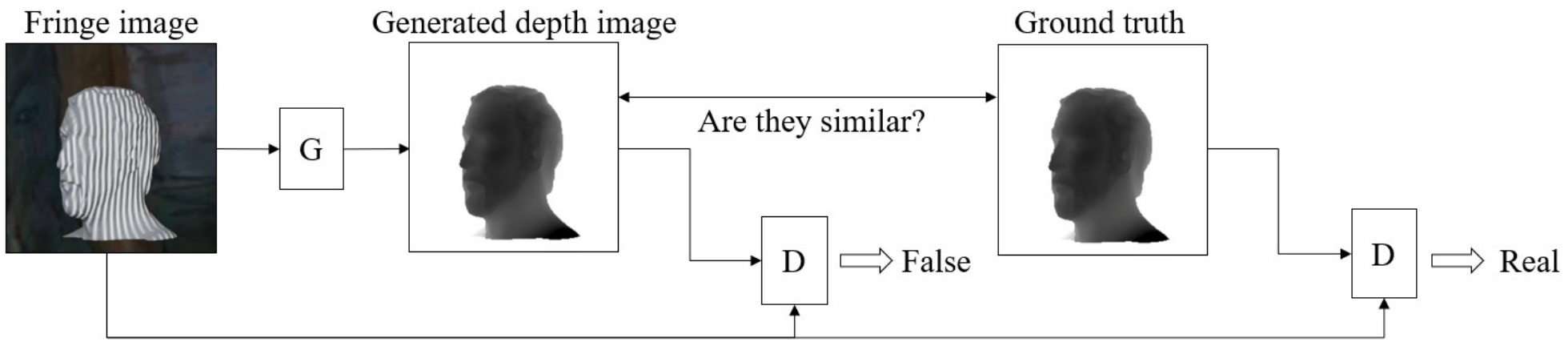

Fig. 6. The Architecture of pix2pix

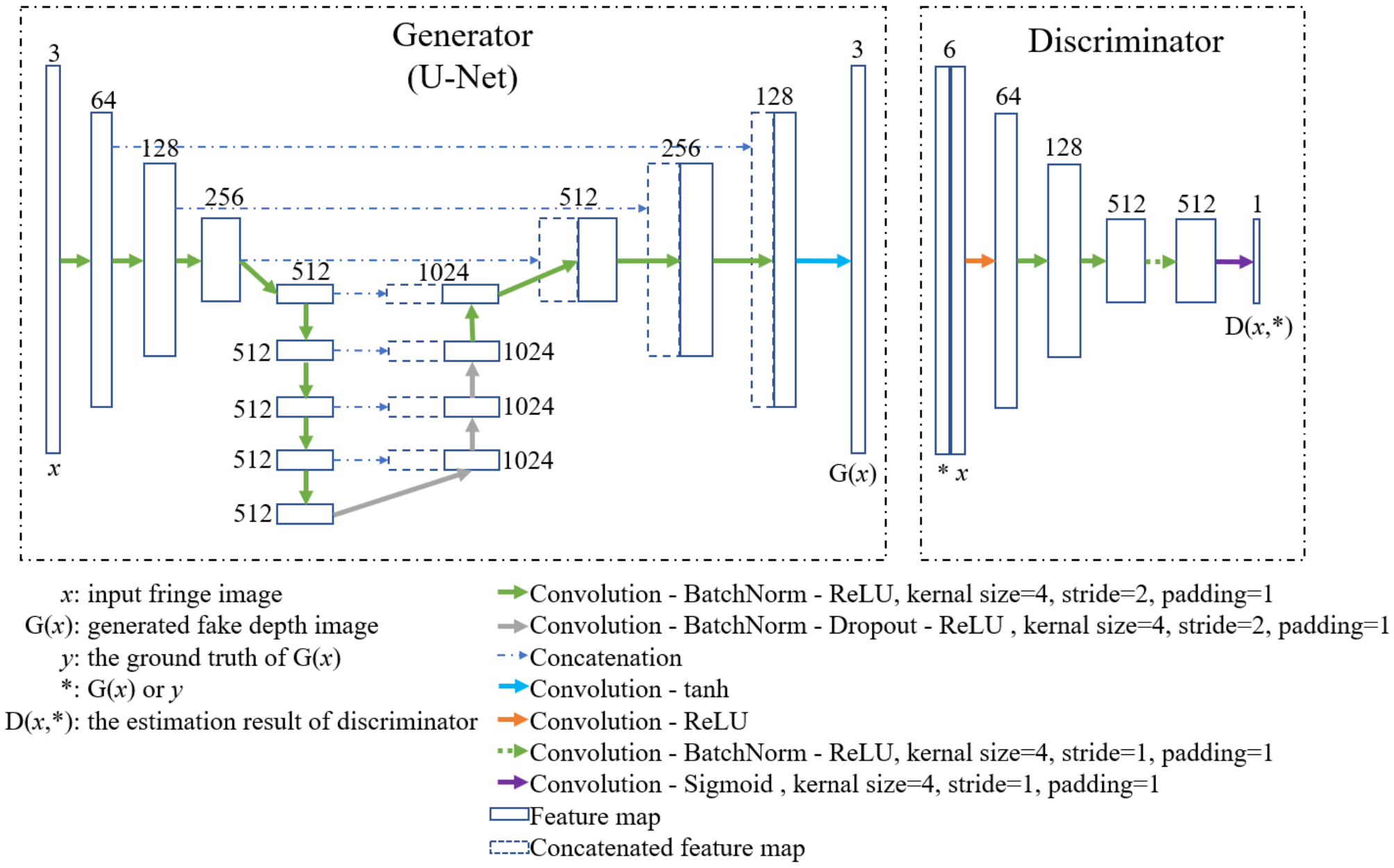

Fig. 7. The network structure of pix2pix, including the structure of U-Net.

### *3.2 The proposed new Loss Function*

A loss function defines the convergence form of a network and so is the key to the quality of outputs. The mean absolute error ($L1$ loss) and the mean square error ($L2$ loss) are the most commonly used loss functions. However, any of the two only evaluates average errors, thus, the resulted outputs quite possibly show low quality in some local regions.

The task in this paper is to retrieve a depth image that records the 3D shape of a scanned object; hence the geometry and the spatial structure is a good constraint to keep the overall effect of the outputs. An index called Structure SIMilarity (SSIM) [43] leverages the structural information to evaluate image quality, which is defined as

$$\mathrm{SSIM}(u,v)=\frac{(2\mu_u\mu_v+c_1)(2\sigma_{uv}+c_2)}{(\mu_u^{\,2}+\mu_v^{\,2}+c_1)(\sigma_u^{\,2}+\sigma_v^{\,2}+c_2)}, \tag{4}$$

where $\mu_u$ is the mean of the evaluated image $u$, $\mu_v$ is the mean of the ground truth $v$, $\sigma_u$ and $\sigma_v$ are the variances of $u$ and $v$, respectively, $\sigma_{uv}$ is the covariance of $u$ and $v$, and $c_1$ and $c_2$ are two constants to avoid division by zero. The SSIM ranges in [0,1], and it is scored low if the evaluated image is compressed, blurred or noise contaminated. With this good ability to measure overall structure of 3D shapes, we take the index SSIM as a term of the loss function. The detailed description of this term is

$$L_{T1}=1-\mathrm{SSIM}(\mathrm{G}(I),d), \tag{5}$$

where $I$ is an input fringe image, G($I$) is the fake depth image generated by U-Net or pix2pix's generator, and $d$ is the ground truth of G($I$).

With the overall accuracy ensured, the local detail is another essential factor to the restoration of a 3D shape. As details are always embedded in the edges of an image, we add another term involving a common tool for edge detection, the Laplacian operator, to the loss function to estimate the detail's similarity between G($I$) and $d$. The added term is described as

$$L_{T2}=\left\|\mathrm{La}(\mathrm{G}(I)),\mathrm{La}(d)\right\|_1, \tag{6}$$

where La(·) denotes convolving an image by the Laplacian operator. With this term added, the network is more sensitive to slight variations of depth, then not only the accuracy of details is improved but also the abrupt artifacts in the generated depth images are eliminated.

Based on the above analysis, for the U-Net, we replace the commonly used $L1$ loss or $L2$ loss with our new loss function below:

$$L_{U-Net}=\lambda_1 L_{T1}+\lambda_2 L_{T2}, \tag{7}$$

where $L_{T1}$ and $L_{T2}$ have been given in Eqs. (5) and (6), respectively, and $\lambda_1$ and $\lambda_2$ are the adjustable weights.

For the pix2pix, we define the loss function as

$$L_{pix2pix}=L_{cGAN}+\lambda_1 L_{T1}+\lambda_2 L_{T2}\,. \tag{8}$$

In Eq.(8), the last two terms are the same to the ones in Eq.(7), and $L_{cGAN}$ is a unique term for cGAN to assess the accuracy of the discriminator's output by

$$L_{cGAN}=\frac{1}{2}\left\|\mathrm{D}(I,\mathrm{G}(I))\right\|_2+\frac{1}{2}\left\|1-\mathrm{D}(I,d)\right\|_2, \tag{9}$$

where D(·) denotes the estimation result of the discriminator that distinguishes between the fake depth image G($I$) and its ground truth $d$ by comparing their relationships to the input fringe image $I$. Note that the $L2$ loss is exploited in Eq.(9) to replace the cross-entropy loss used in original pix2pix[32] since it shows better performance in improving the quality of the result and the stability of the training process [44].

## 4. Experiments

### *4.1 Dataset rendering and data preprocessing*

In this paper, we choose 624 models (some are shown in Fig. 1) from Thingi10K, which covers a rich variety of items with various complexities. To ensure the generalization of the trained model, we separate the 624 models into 13 groups and set different rendering parameters analyzed in section 2.3 for each group to simulate the possible various situations in practice.

The variation range we set for each parameter in Blender is shown in Table 1 and the rendered image pairs are shown in Fig. 8.

**Table 1. The variation range for each parameter**

| parameters | range |
|---|---|
| The period of fringes | [4.4, 6.6] |
| The rotation of fringes (°) | [-5, 5] |
| The angle between the camera and the projector (°) | [10, 20] |
| The power of the projector (W) | [20, 55] |
| The brightness of the ambient light | [0, 1] |
| The rotation of the environment map (°) | [0,360] |

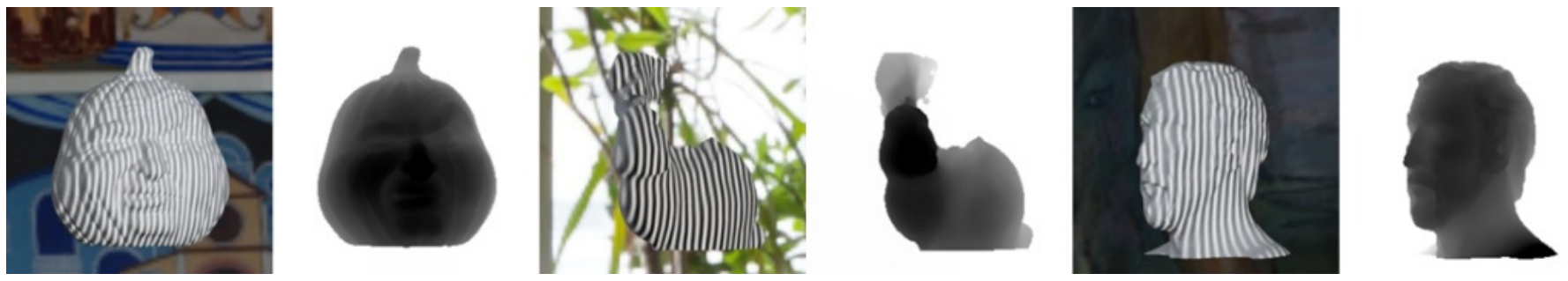

Fig. 8. Some examples of the rendered fringe images and their corresponding depth images.

To enrich the dataset, we render multiple images for each model. In the camera coordinate system shown in Fig. 9, the model is firstly rotated around the $y$-axis by 12 times with each time 30°, and then for each rotation around the $y$-axis, another 12 times of rotation are repeated around the $z$-axis by 5° each time. Therefore, there are 144 fringe images rendered for each object, and a depth image is rendered corresponding to each fringe image. In total, 89856 pairs of images are obtained to create the dataset. We randomly allocate the 624 models to the training and test sets in a ratio of 8.5:1.5, and hence there are no identical objects in the training set and the test set.

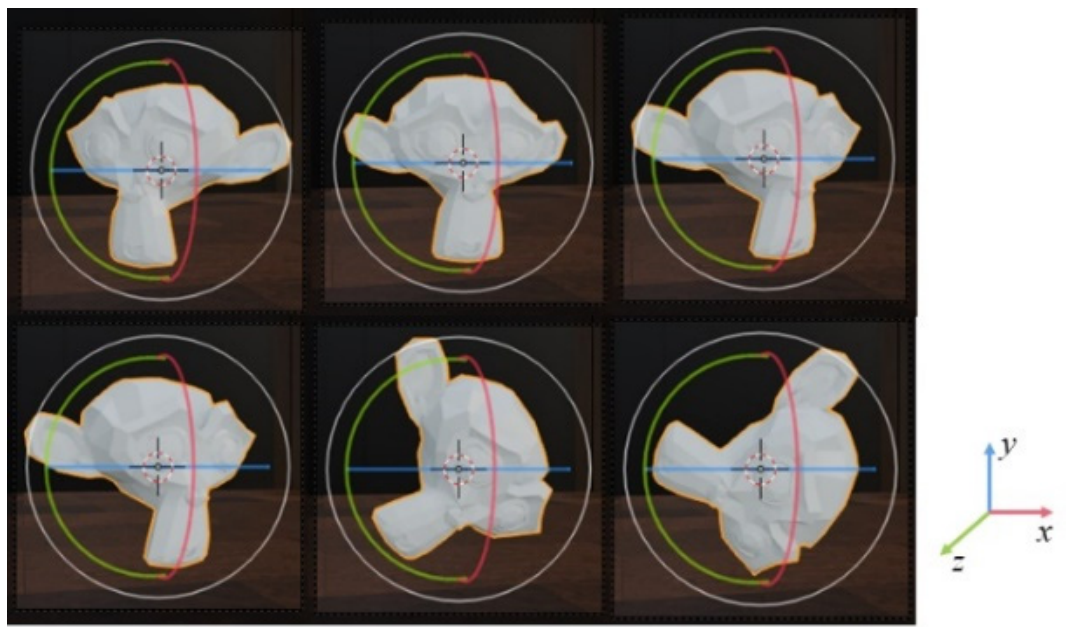

Fig. 9. The rotations of objects. First row: rotation around the $y$-axis of camera coordinate system; second row: rotation around the $z$-axis of camera coordinate system.

Before training, each fringe image and depth image $I$ is normalized to [-1, 1] by

$$I_{nI} = \frac{\max(I) - \min(I)}{I - \min(I)}, \tag{10}$$

and

$$I_{n2} = \frac{I_{n1} - 0.5}{0.5}. \tag{11}$$

### *4.2 Comparison of different loss functions*

We implement ablation experiments to compare the effect of different loss functions and their combinations based on U-Net and pix2pix. During the training process, we use Adam optimizer with momentum parameters $\beta_1$=0.5 and $\beta_2$=0.999. The batch size is 4, with a learning rate of 0.0003. The size of the SSIM window is 8. All the networks with different loss functions are trained for 13 epochs, which is enough for convergence. We set $\lambda_1$ and $\lambda_2$ in Eq.(7) and Eq.(8) as 100 and 10, respectively, which are the best empirical values.

Fig. 10 illustrates the qualitative comparison of different loss functions. All fringe images in Fig. 10 are chosen from the test set, in which the objects have not been seen by the network during training. Fig. 10(a) and (b) represent the rendered fringe images and depth images (ground truth) respectively. Fig. 10(c)-10(e) are the results of U-Net with the proposed loss SSIM+Laplace, the loss with only SSIM term, and only $L1$ loss respectively. Similarly, Fig. 10(f)-10(h) show the corresponding results using pix2pix. No matter for U-Net or for pix2pix, our proposed loss function performs the best in eliminating the artifacts (the red boxes in the first two rows) and keeping details (the red boxes in the last two rows).

To quantify the effect of the results, we further compute the mean absolute error (MAE) and the mean standard deviation of errors (MSDE) for all the results in the test sets, as listed in Table 2. The results in Table 2 accord with Fig. 10 basically. They all prove that our proposed loss function is much effective. Therefore, we adopt SSIM+Laplace as the loss function of the following experiments. Fig. 10 and Table 2 also show that U-Net generally performs better than pix2pix both in quality and in quantity.

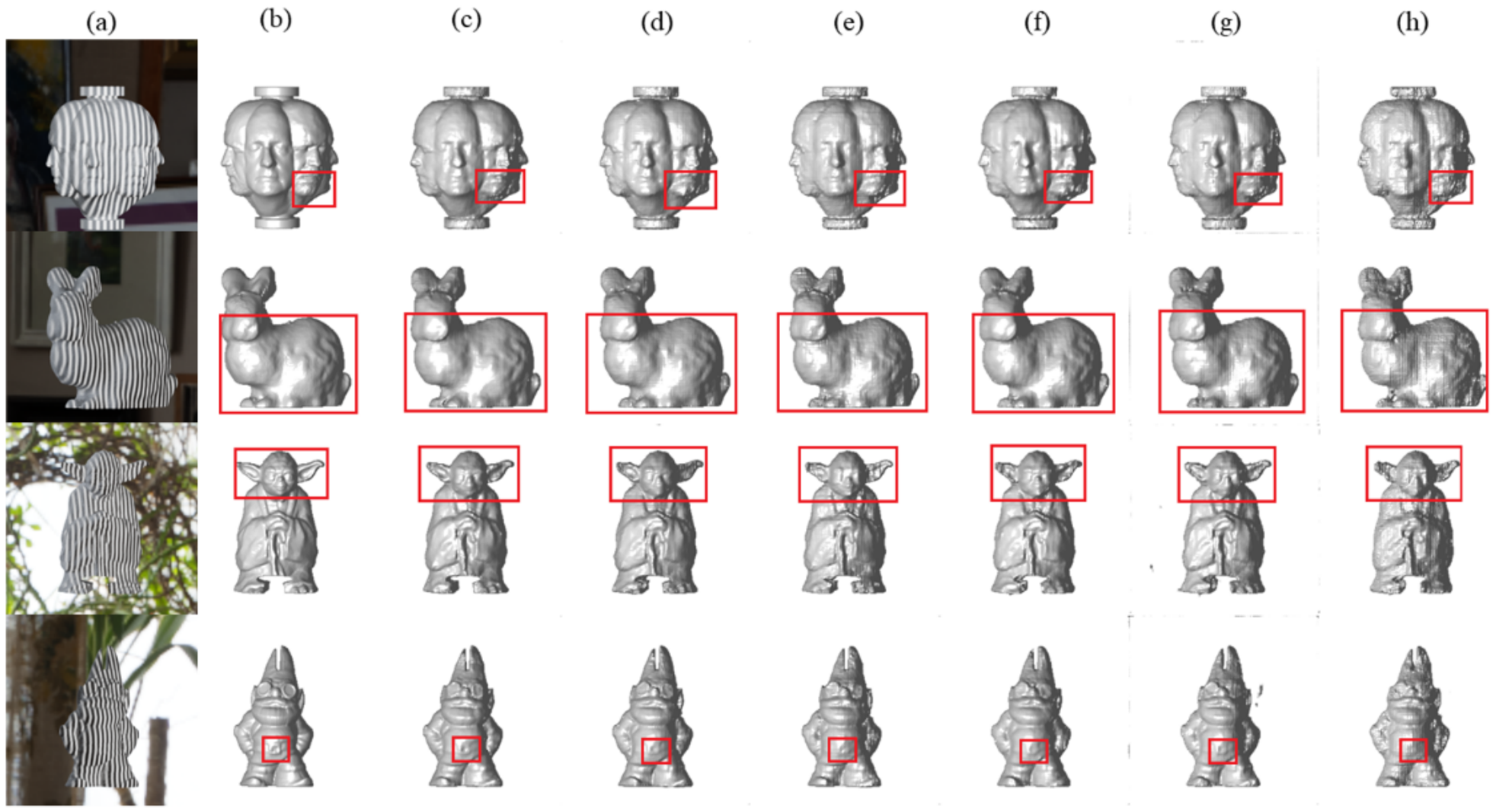

Fig. 10. The ablation experiments of different output depth images result from different losses. (a) Fringe image, (b) Ground truth, (c) U-Net with SSIM+Laplace, (d) U-Net with SSIM, (e) U-Net with L1, (f) pix2pix with SSIM+Laplace, (g) pix2pix with SSIM, (h) pix2pix with L1.

**Table 2. The quantitative metrics of different loss functions**

| | | MAE | MSDE |
|---|---|---|---|
| U-Net | L1 | 0.0257 | 0.0762 |
| | SSIM | 0.0263 | 0.0713 |
| | SSIM+Laplace | 0.0260 | 0.0708 |
| pix2pix | L1 | 0.0300 | 0.0974 |
| | SSIM | 0.0280 | 0.0813 |
| | SSIM+Laplace | 0.0281 | 0.0807 |

### *4.3 The impact of improving system generalization ability on accuracy*

In this section, we explore the impact of different rendering parameters on the accuracy of the depth images generated by the network. With the models separated into 13 groups, the datasets are rendered in three different cases:

$D_1$) the parameters in Table 1 are all the same for rendering all groups of images;

$D_2$) the first three parameters in Table 1 are different from group to group;

$D_3$) all the parameters in Table 1 vary among different groups.

Appendix B gives the other common settings of $D_1$, $D_2$, and $D_3$. Each dataset above is divided into a training set and a test set by 8.5:1.5 and then U-Net and pix2pix are trained by $D_1$, $D_2$, and $D_3$ separately. Table 3 records the MAE and MSDE of the test set in each dataset. It shows that for both U-Net and pix2pix, the accuracy of the test set decreases as the complexity of the dataset increases. Fig. 11 illustrates the depth images generated from a real captured fringe image by U-Net and pix2pix trained by $D_1$, $D_2$, and $D_3$, respectively. The fringe image in Fig. 11 is captured by an arbitrary FPP system under arbitrary indoor lighting, and the object "Venus" has not appeared in any training datasets. It is obvious that the results of $D_3$ are the best because it considers the most disturbances of the real FPP system. Therefore, the generalization ability of the network shows much better if more system variables and environment variables are considered in rendering the dataset.

**Table 3. The MAE and SDE of the test set in each dataset**

| | | MAE | MSDE |
|---|---|---|---|
| U-Net | $D_1$ | 0.0095 | 0.0381 |
| | $D_2$ | 0.0197 | 0.0525 |
| | $D_3$ | 0.0260 | 0.0708 |
| pix2pix | $D_1$ | 0.0102 | 0.0412 |
| | $D_2$ | 0.0209 | 0.0554 |
| | $D_3$ | 0.0281 | 0.0807 |

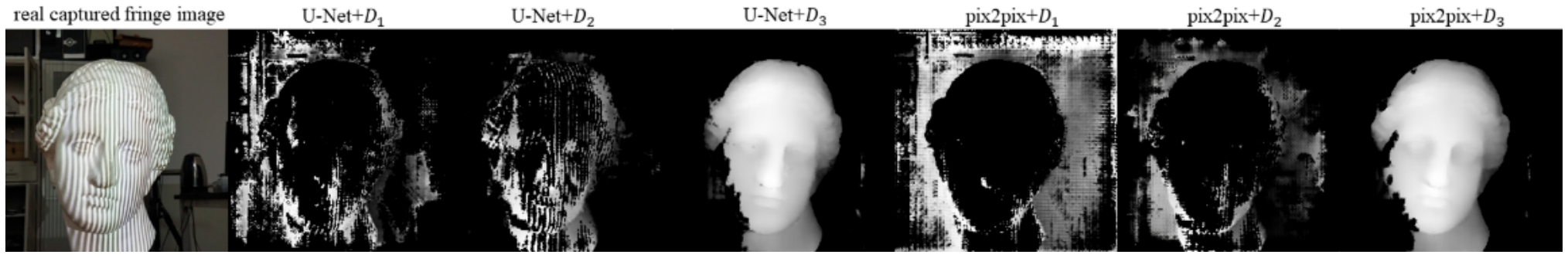

Fig. 11. The depth images generated from real captured fringe image by $D_1$, $D_2$ and $D_3$.

### *4.4 The practicability of the network*

In this section, we test the practicability of our trained network by some fringe images captured from different real FFP systems that are casually placed in different scenarios. The tested objects are randomly selected from common items and have not appeared in the training dataset.

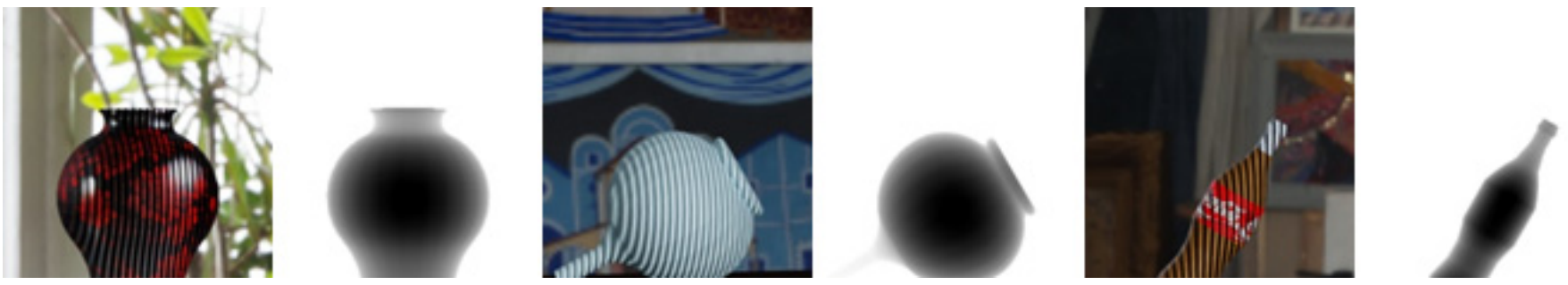

Fig. 12. The examples of the fringe images and their corresponding depth images generated by the models in ShapeNet.

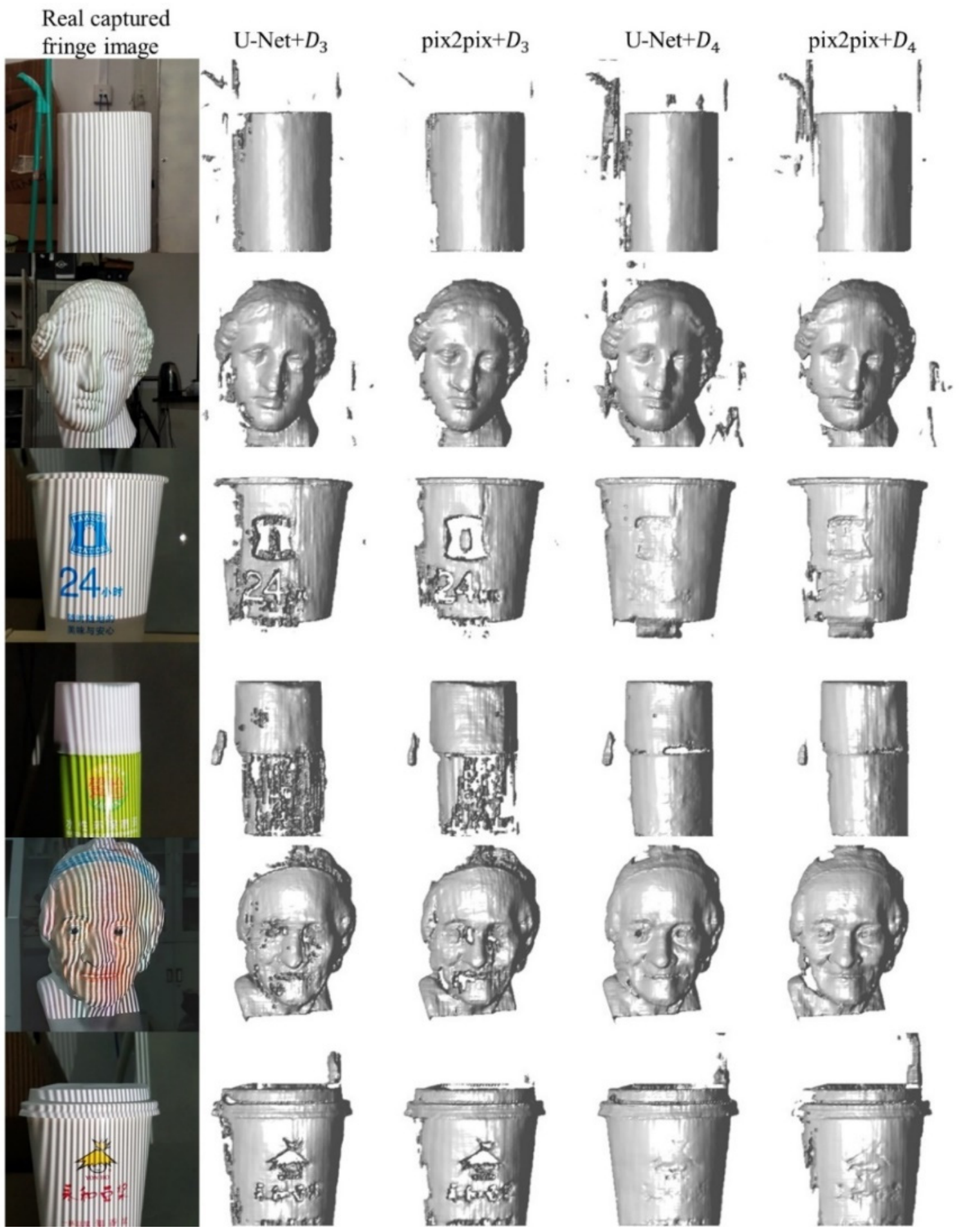

Fig. 13. Different output depth images result from Different datasets.

To test the objects with colorful surfaces, the dataset $D_3$ is further extended to form a new dataset $D_4$, by adding some images rendered by 3D models in ShapeNet [37], which have colorful surfaces and textures. The newly selected 320 models in ShapeNet are divided into 8 groups and each group is also rendered by setting with different parameters in Table 1. Then each model is rotated 144 times, and 46,080 new image pairs are rendered, as exampled in Fig. 12. The new dataset $D_4$ is also divided into a training set and a test set by 8.5:1.5.

The test results are shown in Fig. 13, which illustrate that the network has good potentials in practical applications. By comparison, the 3D shapes generated by pix2pix present more obvious strip-like artifacts, illustrating that the U-Net still performs better than pix2pix.

## 5. Conclusion and discussion

In this paper, we propose a method to construct a virtual FPP system with computer graphics, and we also analyze the key factors being able to be set in the virtual FPP systems to produce samples much close to the practical. This virtual system provides great convenience for the data collection of deep learning for FPP. To enhance the accuracy of the restored 3D shape, we also propose an effective new loss function combining the SSIM index and Laplace operator. Abundant simulation experiments and real experiments are conducted based on the most powerful and representative networks, U-Net and pix2pix. The results show that the networks trained by the proposed loss function and the dataset generated by our virtual system have satisfied accuracy and generalization.

The comparisons illustrate that U-Net performs better than pix2pix with relatively sufficient samples. The main reason is that U-Net is adept at extracting a certain mapping relationship, such as the mapping from fringe to depth in this work. In fact, pix2pix shows better than U-Net if training samples are reduced largely in our other experiments. To explore the method of accurately extracting depth with fewer samples is worthy of deeper studies. In addition, to retain more accurate details and to convert the depth to a 3D point cloud are all the future works.

## Acknowledgements

This work is supported by the National Natural Science Foundation of P. R. China (61828501), the Special Project on Basic Research of Frontier Leading Technology of Jiangsu Province of China (Grant Nos. BK20192004C) and Natural Science Foundation of Jiangsu Province of China (Grant Nos. BK20181269).

## Appendix A

The following explains the shading tree nodes of the projector in Blender:

1) Geometry-normal: "normal" refers to the vector pointing from the projector to a certain point on the surface of the object.
2) Mapping-Point: the rotation (X/Y/Z) here decides the direction of the emitting light.
3) SeparateXYZ-Divide-CombineXYZ: project the 3D vector to the *xy*-plane, so that the projected pattern is only related to the *x* and *y* coordinates, not to the *z* coordinate.
4) The second Mapping-point: change the position, direction, and size of the projection pattern. Because the origin of the projection pattern is in the upper left corner, the "X" and "Y" coordinates of the "Location" are offset by 0.5 meters.
5) sin0.bmp: set the projection pattern (fringe image).
6) Light Falloff: set the way the light intensity decreases with distance.
   a) Strength: light intensity before applying attenuation (Light Falloff node).
   b) Constant: set a constant light attenuation.
7) Emission: add Lambertian luminous shader for light output.
8) Light Output: light output.

## Appendix B

**Common settings:**
Camera mode: Perspective
Camera field of view: 7°
Projector size: 0.001m
Position of the background wall when rendering depth image: (0, 0.05m, 0)
Position of the 3D model: (0, 0, -0.02m)
Position the projector: (0, -1.5m, 0)

Size of the 3D model: the maximal dimension is scaled to 0.14m.

**Note:**

When rendering depth images, keep the positions of the camera and the object unchanged, and import a plane behind the object, otherwise the depth image would record the depth of the background regions as infinite.